\documentclass[aps,prl,preprint,superscriptaddress]{revtex4}

\usepackage{epsfig} 

\begin{document}

\title{Defective vortex lattices in layered superconductors with 
point pins at the extreme type-II limit}
\author{J. P. Rodriguez}
\affiliation{Dept. of Physics and Astronomy, California State University,
Los Angeles, California 90032.}

\date{\today}

\begin{abstract}
The mixed phase of layered superconductors 
with no magnetic screening
is studied through a partial duality analysis of the corresponding
frustrated $XY$ model in the presence of weak random point pins.
Isolated layers exhibit a defective vortex lattice at low temperature
that is phase coherent.
Sufficiently weak Josephson coupling between adjacent layers
results in an entangled vortex solid that exhibits
weak superconductivity across layers.
The corresponding vortex liquid state shows an inverted specific heat
anomaly that we propose accounts for that seen in YBCO.
 A three-dimensional
 vortex lattice with dislocations occurs at stronger coupling.
This crossover sheds light on the apparent discrepancy
concerning the observation of a vortex-glass phase
in recent Monte Carlo simulations of the same $XY$ model.
\end{abstract}

\maketitle

High-temperature superconductors are 
layered and extremely type-II\cite{blatter}.
This fact led  to the suggestion early on that a state 
with phase-coherent yet decoupled layers
is  possible\cite{friedel}.
It was  later demonstrated, however, that {\it any} amount of Josephson
coupling between layers results in a macroscopic Josephson effect across layers
at low temperature\cite{korshunov}.
In the presence of external magnetic field perpendicular
to the layers,  other workers made the analogous proposal
that  sufficiently
weak coupling could lead to a  decoupled stack of phase-coherent
two-dimensional (2D) vortex lattices\cite{supersolid}\cite{ffh}.
Monte Carlo simulations\cite{koshelev97} and  a partial duality
analysis\cite{jpr02}\cite{cec03}
 of the corresponding frustrated $XY$ model
demonstrate, however, that a highly entangled vortex lattice
state with  relatively small or no  phase coherence across 
layers does not exist in practice.

The elusive decoupled vortex lattice state may
exist at low  temperature in the presence
of  random point pins, however\cite{larkin}\cite{bulaevskii}.
In this paper,  we show that this is indeed the case through
a partial duality analysis of the corresponding layered $XY$ model
with uniform frustration\cite{jpr00}.
We show first that  a defective vortex lattice state\cite{ffh}  can
 exist in isolated layers.
It 
exhibits macroscopic phase coherence in the zero-temperature limit,
 despite the presence of unbound
dislocations that are assumed to be quenched in by
the random pins\cite{n-s}.
We next turn on  Josephson coupling between adjacent
layers and  find that 
weak superconductivity exists across layers at sufficiently
high layer anisotropy\cite{jpr00}    in the zero-temperature limit.
After assuming a continous 2D ordering transition for each layer in
isolation, we then find that an inverted specific heat jump can occur
inside of  the vortex liquid state at weak coupling.
This prediction compares favorably with the
recent observations of such a peak
in the high-temperature superconductor YBCO\cite{bouquet}.

{\it 2D.}
Consider a stack of isolated superconducting layers 
in a perpendicular external magnetic field.
In the absence of Josephson coupling 
as well as of magnetic screening, 
the $XY$ model over the square lattice with
uniform frustration provides a qualitatively
correct description of the mixed phase in each layer.
The corresponding Boltzmann distribution is set by the
sum of energy functionals
\begin{equation}
  E_{XY}^{(2)}(l) = - \sum_{\mu = x, y} \sum_{\vec r} J_{\mu} \,
{\rm cos} [\Delta_{\mu} \phi  - A_{\mu}]
\label{2DXY}
\end{equation}
for the superfluid kinetic energy of each layer $l$
written in terms of the superconducting phase 
$\phi (\vec r, l)$.
Here $\Delta_{\mu} \phi (\vec r, l) =
\phi(\vec r + a \hat\mu, l) - \phi(\vec r, l)$ and
$\vec A = (0, 2\pi f x/ a)$ make up the local supercurrent,
where $f$ denotes the concentration of vortices over the
square lattice, with lattice constant $a$.
The local phase rigidity $J_{\mu} (\vec r, l)$
is assumed to be constant over most of the 
nearest-neighbor links $(\vec r, \vec r + a \hat\mu )$
in layer $l$,
with the exception of those links in the vicinity of
the  pinning sites that are located at random.
After taking  the Villain approximation,
which is generally valid at low temperature \cite{villain},
a series of standard manipulations then lead to  
a Coulomb gas ensemble with pins    that
describes   the vortex degrees of 
freedom on the dual square lattice\cite{cec03}. 
The ensemble for each layer $l$ is weighted by the Boltzmann
distribution set by the energy functional
%
\begin{equation}
  E_{\rm vx} (l) =  (2\pi)^2   \sum_{(\vec R, \vec R^{\prime})}
\delta Q\ J_0 G^{(2)} 
\ \delta Q^{\prime}
+  \sum_{\vec R} V_p  \, | Q  |^2
 \ ,
\label{evx}
\end{equation}
%
written in terms of the integer vorticity field $Q (\vec R, l)$ over
the sites $\vec R$ of the dual lattice in that  layer,
and of the fluctuation $\delta Q = Q - f$.
A logarithmic interaction, $G^{(2)} = - \nabla^{-2}$,
 exists between the vortices, with a strength $J_0$ equal
to the gaussian  phase rigidity.
Last, $V_p (\vec R, l)$
is the resulting pinning potential\cite{cec03}.

We shall next assume that the array of random pins
in each layer, $V_p(\vec R,l)$,
quenches in unbound dislocations into the triangular vortex lattice 
at zero temperature\cite{n-s}.  
To check for superconductivity in such a defective 2D vortex lattice, we
now  compute the macroscopic phase rigidity,
which is  given by
one over the dielectric constant of the 2D Coulomb gas (\ref{evx}) \cite{eps}:
\begin{equation}
{\rho_s^{(2D)}  /    {J_0}}    = 
 1 - \lim_{k \rightarrow 0} (2\pi / \eta_{\rm sw})
\langle \delta Q_{\vec k} \delta Q_{-\vec k} \rangle
 / k^2 a^2 {\cal N_{\parallel}}  \ .
\label{epsinv}
\end{equation}
Here  $\delta Q_{\vec k} =  Q_{\vec k} - \langle Q_{\vec k}\rangle$
is the fluctuation in the Fourier transform of the vorticity in layer $l$: 
$Q_{\vec k} = \sum_{\vec R} Q (\vec R, l) e^{i \vec k\cdot \vec R}$.
 Also,
$\eta_{\rm sw} = k_B T / 2\pi J_0$ 
is the  spin-wave component of the
phase-correlation exponent,
and 
$\cal N_{\parallel}$ denotes the number of points in the square-lattice grid.
Now suppose that a given vortex is displaced
by $\delta\vec u$ with respect to its location at zero temperature.
Conservation of vorticity dictates that 
its 
fluctuation is given by
$ - \vec\nabla\cdot\delta\vec u$.
Substitution
into Eq. (\ref{epsinv}) then yields the result
%
\begin{equation}
\rho_s^{(2D)} / J_0 = 1 - ( \eta_{\rm vx}^{\prime} / \eta_{\rm sw} )
\label{rho2d}
\end{equation}
for the phase rigidity in terms of the
vortex component to the phase-correlation exponent,
\begin{equation}
\eta_{\rm vx}^{\prime} = \pi \Bigl\langle
\Bigl[\sum_{\vec R}^{\qquad\prime} \delta\vec u \Bigr]^2\Bigr\rangle /
N_{\rm vx} a_{\rm vx}^2.
\label{etavx1}
\end{equation}
The latter monitors fluctuations of the
center of mass of the vortex lattice\cite{jpr01}.
Above, $N_{\rm vx}$ denotes the number of vortices, 
while $a_{\rm vx} = a / f^{1/2}$.

To proceed further,
we now express
the displacement field  as a superposition of {\it pure}
wave and defect components of the 
 triangular
vortex lattice\cite{jpr01}:
$\delta\vec u = \delta\vec u_{\rm wv} + \delta\vec u_{\rm df}$.
Notice by Eq. (\ref{etavx1})
 that phase coherence is insensitive to the wave contribution
if rigid translations are excluded,
since $\sum \delta\vec u_{\rm wv} = 0$ in such case.
The former is achieved through bulk pinning\cite{cec03},
and the latter then follows under periodic boundary conditions.
Consider now a single unbound  dislocation
with Burgers vector $\vec b$
that slides along it glide plane\cite{blatter}
a distance $\delta R_{\rm df}$ with respect to its location
at zero temperature.
The relative displacement field, $\delta\vec u_{\rm df}$, then
corresponds to that of a {\it pure} dislocation pair 
of extent $\delta R_{\rm df}$ that is  oriented along its glide plane.
After following steps similar to those taken
in ref. \cite{jpr01} for the pristine case, 
it can be shown that  Eq. (\ref{etavx1})
yields a fluctuation of the center of mass
\begin{equation}
\eta_{\rm vx}^{\prime}   \cong
   n_{\rm df}
\overline {\langle |\delta  R_{\rm df} |^2 \rangle}
(b / 2 a_{\rm vx})^2 {\rm ln}\, R_0 / a_{\rm df} 
\label{etavx2}
\end{equation}
for the vortex solid,
where $n_{\rm df}$ denotes the density of unbound dislocations, where
 $a_{\rm df}$ is the core diameter of a   dislocation, and where 
$R_0$ is an infrared cut-off.  Above, the overbar 
denotes a bulk  average.
Observe now that 
both $\eta_{\rm sw}$ {\it and} $\eta_{\rm vx}^{\prime}$ vanish linearly
with temperature.
By Eq. (\ref{rho2d}), we conclude that the
defective 2D vortex lattice shows a positive phase rigidity
in the zero-temperature limit at sufficiently
dilute concentrations of unbound dislocations, $n_{\rm df}\rightarrow 0$.
The above is borne out by direct Monte Carlo simulations\cite{mc}
of the 2D Coulomb gas ensemble (\ref{evx}). 

The previous positive result for macroscopic phase coherence
[Eq. (\ref{rho2d})] in the zero-temperature limit
can be confirmed by calculation
of generalized phase auto-correlation functions
 within  an isolated layer:
$C_l [q] = \langle  {\rm exp} [ i\sum_{\vec r} q(\vec r)
\cdot \phi (\vec r, l)] \rangle_{0}$.
Following a similar calculation in the pristine case\cite{jpr01},
application of the Villain approximation 
[see  Eq. (\ref{evx}) and ref. \cite{villain}]
yields the form
$C_l [q] = |C_{l}[q]| {\rm exp} [i\sum_{\vec r} q(\vec r)  \phi_0(\vec r, l)]$
for these autocorrelations,
where $\phi_0 (\vec r, l)$ represent the zero-temperature configurations
of isolated layers.
In the low-temperature regime,
phase correlations  are then found to  decay algebraicly  like
\begin{equation}
|C_l [q]|
    = g_0^{n_+}\cdot {\rm exp}\Bigl[  \eta_{2D}
\sum_{(1,2)} q(1){\rm ln} (r_{12} / r_0)\, q(2)\Bigr]
\label{clq}
\end{equation}
at the asymptotic limit, $r_{12}\rightarrow\infty$,
with a net correlation exponent approximately equal to
$\eta_{2D} \cong  \eta_{\rm sw} + \eta_{\rm vx}^{\prime}$
for small vortex components, $\eta_{\rm vx}^{\prime}\ll \eta_{\rm sw}$.
Here, 
$g_0 = \rho_s^{(2D)} / J$ is the ratio of the 2D stiffness
with its value at zero temperature, $J$,
while  $n_+$ counts half the number of probes in $q (\vec r)$.
Also, $r_0$ denotes the natural ultraviolet scale.

{\it 3D.}
We shall now add     a weak Josephson coupling energy
$-J_z {\rm cos} (\Delta_z \phi - A_z)$
to all of the verticle links
in between adjacent layers of the  three-dimensional (3D) $XY$ model.
Here, 
$J_z = J/\gamma^{\prime 2}$ is the perpendicular coupling
constant, with anisotropy parameter $\gamma^{\prime} > 1$.
The layered $XY$ model can be effectively analyzed in
the selective high-temperature limit, $k_B T\gg J_z$,
through  a {\it partial} duality transformation.
This leads to a dilute Coulomb gas (CG) ensemble
 that describes the nature of the Josephson coupling in terms
of  dual charges that live on the vertical links.
Phase correlations across  layers 
can then be computed
from the quotient
\begin{equation}
  \Bigl\langle {\rm exp} \Bigl[i\sum_r p(r) \phi(r)\Bigr]\Bigr\rangle =
Z_{\rm CG}[p]/Z_{\rm CG}[0]
\label{quo}
\end{equation}
of partition functions for
a    layered CG ensemble\cite{jpr00}:
\begin{equation}
  Z_{\rm CG}[p] = \sum_{\{n_{z}(r)\}} y_0^{N[n_z]}
\Pi_{l} C_l [q_l]
\cdot e^{-i\sum_r n_z A_z},
\label{z_cg}
\end{equation}
where the dual charge, $n_z (\vec r, l)$, is an integer field
that lives on links between adjacent layers $l$ and $l+1$
located  at 2D points $\vec r$.
The ensemble is weighted
by a product
of phase auto-correlation functions
for isolated layers $l$
probed at the dual   charge  that accumulates onto
that layer:
\begin{equation}
 q_l (\vec r) = p(\vec r, l) +  n_z (\vec r, l-1) - n_z (\vec r, l).
\label{ql}
\end{equation}
It is also weighted
by a bare fugacity
$y_0$   that is
raised to the power
$N [n_z]$
equal to the total
 number of dual charges, $n_z = \pm 1$.
The fugacity is
given by
$y_0 = J_z / 2 k_B T$ in the selective high-temperature regime,
$J_z \ll k_B T$,
reached at large model  anisotropy.

In the absence of Josephson coupling,
random point pins lead to 
zero-temperature phase configurations,
$\phi_0 (\vec r, l)$, 
that are completely uncorrelated across layers. 
At zero parallel field,  Eqs. (\ref{quo}) and (\ref{z_cg})
therefore yield the expressions\cite{koshelev96}
\begin{equation}
\overline{\langle {\rm cos}\, \phi_{l, l+1}\rangle} \cong 
y_0 \sum_1 \overline{C_l (0,1)} \cdot \overline {C_{l+1}^* (0,1)}
\label{cos1}
\end{equation}
and
\begin{equation}
\overline{|\langle e^{i\phi_{l, l+1}}\rangle|^2} \cong 
y_0^2 \sum_{1} 
\sum_{2} 
\overline{C_l (0,1) C_l^* (0, 2)} 
\cdot \overline{C_{l+1}(0,2) C_{l+1}^*  (0,1)}
\label{eiphi}
\end{equation}
for the inter-layer ``cosine'' and the
 inter-layer phase correlation, to lowest order in the fugacity.
The overbar represents a  bulk (disorder)  average,
while 
$\phi_{l, l+1} (\vec r) =  
\phi (\vec r, l + 1) - \phi (\vec r, l) - A_z (\vec r, l)$
is the gauge-invariant phase difference across adjacent layers.
Macroscopic phase coherence shown by each layer in isolation
(\ref{rho2d})
is lost at a transition temperature\cite{mc} $T_g^{(2D)} > 0$.
Only short-range phase correlations
on the scale of  $\xi_{2D}$
exist at higher temperature following
$C_l (1,2) = g_0 e^{-r_{12}/\xi_{2D}} e^{i\phi_0 (1)}
e^{-i\phi_0 (2)}$.
By analogy with 2D melting physics\cite{jpr01}\cite{NH},
the presence of quenched-in unbound dislocations also implies  
that 
only short-range phase  correlations
exist inside of each layer in isolation,  on average,  at zero temperature.
Specifically, we have
$\overline{{\rm exp}[i\phi_{l, l+1}^{(0)} (1)]\cdot 
{\rm exp}[-i\phi_{l, l+1}^{(0)} (2)]}
= e^{-2r_{12}/l_{2D}}$
asymptotically, where
$\phi_{l, l+1}^{(0)} (\vec r) =
    \phi_0 (\vec r, l+1) - \phi_0(\vec r, l) - A_z (\vec r, l)$
is the quenched interlayer phase difference, and
where  $l_{2D}$ represents 
a zero-temperature
disorder  scale set by $n_{\rm df}$. 
Substitution into expression (\ref{eiphi}) then  yields the result
$\overline{|\langle e^{i\phi_{l, l+1}}\rangle|^2} \sim
[g_0^2 (J / k_B T) (l_{2D} \xi_{2D} / \Lambda_0^2)]^2$
for the inter-layer phase correlation inside of the critical regime,
$\xi_{2D}\gg l_{2D}$,
where $\Lambda_0 = \gamma^{\prime} a$ is the Josephson penetration length.
This approximate result reaches unity at a cross-over field
\begin{equation}
f \gamma_{\times}^{\prime 2} \sim
g_0^2 (J/k_B T) (l_{2D} \xi_{2D}/a_{\rm vx}^2),
\label{xover}
\end{equation}
in units of the naive  decoupling scale $\Phi_0/\Lambda_0^2$,
that separates 2D from 3D vortex-liquid behavior\cite{jpr02}.
Substitution into expression (\ref{cos1}) for the 
inter-layer ``cosine'',
on the other hand,
yields a  non-divergent result
\begin{equation}
\overline{\langle {\rm cos}\, \phi_{l, l+1}\rangle} \sim
g_0^2 (J / k_B T)   [(l_{2D}^{-1} + \xi_{2D}^{-1})^{-1} / \Lambda_0]^2
\label{cos2}
\end{equation}
that is valid in the decoupled vortex liquid that exists at
fields much larger than $f\gamma_{\times}^{\prime 2}$. 
It can be shown\cite{jpr04} that the next-leading-order term for the 
inter-layer ``cosine''
(\ref{cos1}) is negative, that it  diverges just like the
leading order term 
for the inter-layer correlation (\ref{eiphi}),
and that  it becomes comparable to its own leading
order term precisely at fields below the 2D-3D cross-over scale,
Eq. (\ref{xover}).
Last, Eq. (\ref{cos2})
implies an anomalous inter-layer contribution to the specific heat per volume
 equal to
\begin{equation}
\delta c_{v}^{\perp} 
\cong
 2 [ 1 + (\xi_{2D} /  l_{2D})]^{-1} 
(\partial {\rm ln}\, \xi_{2D}^{-1} /  \partial T ) e_{J},
\label{c_perp}
\end{equation}
where
$e_{J} = \overline{\langle {\rm cos}\, \phi_{l, l+1}\rangle}
\cdot J/\Lambda_0^2 d$
is the 
Josephson energy density,
and where  $d$ denotes the spacing in between adjacent layers.
It also notably shows an inverted
specific heat jump that is followed by a  tailoff
at a temperature $T_p$ such that
$\xi_{2D} (T_p) \sim  l_{2D}$
if $\xi_{2D}$ diverges faster 
than $(T - T_g^{(2D)})^{-1}$.
This approximate result is again valid at high anisotropy, 
$\gamma^{\prime} > \gamma_{\times}^{\prime}$,
which yields  the bound
$l_{2D}  <  g_0^{-1} (k_B T / J)^{1/2} \Lambda_0$ 
on the 2D  disorder scale by Eq. (\ref{xover}).

The previous analysis clearly demonstrates that a selective high-temperature
 expansion in powers of the fugacity $y_0$ necessarily
 breaks down in the ordered phase, $T < T_g^{(2D)}$, 
where $\xi_{2D}$ is infinite. 
At this stage it becomes useful to re-express the layered CG enesemble
(\ref{z_cg}) by replacing
$C_l [q]$	with its magnitude (\ref{clq}), and by compensating
this change with the additional replacement of
$A_z (\vec r, l)$ with
$-\phi_{l, l+1}^{(0)} (\vec r)$.
A Hubbard-Stratonovich 
transformation
of the CG partition function (\ref{z_cg})
reveals\cite{jpr97} that it is 
equivalent  to
 a   renormalized Lawrence-Doniach (LD) model
with an  energy functional that is given by\cite{jpr00}
\begin{equation}
E_{\rm LD} = 
\rho_s^{(2D)}\int d^2 r
\sum_{l} \Biggl[
{1\over 2}(\vec\nabla\theta_l)^2
-\Lambda_0^{-2}
{\rm cos}\, \theta_{l, l+1} \Biggr],
\label{e_ld}
\end{equation}
where 
$\theta_{l, l+1}  = \phi_{l, l+1}^{(0)} +  \theta_{l+1} - \theta_l$.
A standard thermodynamic analysis\cite{jpr02} 
 then yields that
 the strength of the local Josephson coupling is
given by
$\overline{\langle {\rm cos}\, \phi_{l,l+1}\rangle}  = y_0
 +  g_0 \overline{\langle {\rm cos}\, \theta_{l, l+1} \rangle}$.
It can also be shown\cite{jpr00}
 that phase coherence exists across a macroscopic
number of layers, with a corresponding 
phase rigidity equal to
$\rho_s^{\perp} / J_{z} \cong
g_0 \overline{\langle {\rm cos}\, \theta_{l, l+1} \rangle}$.

In order to compute $\overline{\langle {\rm cos}\, \theta_{l, l+1} \rangle}$
at low  temperature, we must first determine
the configuration
that optimizes $E_{\rm LD}$.  Eq. (\ref{e_ld}) implies that it satisfies the
field equation 
\begin{equation}
-\nabla^2 \theta_l^{(0)} +\Lambda_0^{-2}[{\rm sin}\, \theta_{l-1, l}^{(0)}
- {\rm sin}\, \theta_{l, l+1}^{(0)}] = 0.
\label{fldeqs}
\end{equation}
In the weak-coupling limit,  $\Lambda_0\rightarrow\infty$,
we therefore have that 
$\theta_l^{(0)} (\vec r)$  is constant inside of each layer.
The fact that $\overline{e^{i\phi_0 (1)} e^{-i\phi_0(\infty)}} = 0$ 
then implies that
$\overline{{\rm cos}\, \theta_{l, l+1}} = 0$ at zero temperature in the 
weak-coupling limit.
Indeed, the LD ``cosine'' can be calculated
perturbatively, where one finds that
$\overline{{\rm cos}\, \theta_{l, l+1}} 
\sim  ( l_{2D}/\Lambda_0)^2 {\rm ln}(\Lambda_0 / l_{2D})$
at  zero temperature\cite{bulaevskii}\cite{jpr04}. 
In the opposite limit of weak disorder, $l_{2D}\rightarrow\infty$,
Eq. (\ref{fldeqs}) yields that
${\rm sin}\, \theta_{l-1, l}^{(0)}
= {\rm sin}\, \theta_{l, l+1}^{(0)}$,
on the other hand.
This then implies that
${\rm cos}\, \theta_{l, l+1}^{(0)} = 1$
in the weak disorder limit.  The bulk average
$\overline{{\rm cos}\, \theta_{l, l+1}}$ at zero temperature
must therefore pass
between zero and unity
at   $\Lambda_0\sim l_{2D}$.  
This condition  defines a decoupling cross-over field
$f\gamma_D^{\prime 2}(0)\sim ( l_{2D}/a_{\rm vx})^2$
in units of $\Phi_0/\Lambda_0^2$,
at which point the reversible magnetization shows a broad
diamagnetic peak.
By the discussion following Eq. (\ref{e_ld}),
we conclude that
random point pins result in a vortex glass at sufficiently
high layer  anisotropy\cite{larkin}\cite{bulaevskii},
$\Lambda_0\gg l_{2D}$, 
that exibits weak superconductivity
across layers\cite{jpr00}: 
$\rho_s^{\perp} \ll J_z$.

\begin{figure}
\includegraphics[scale=0.36, angle=-90]{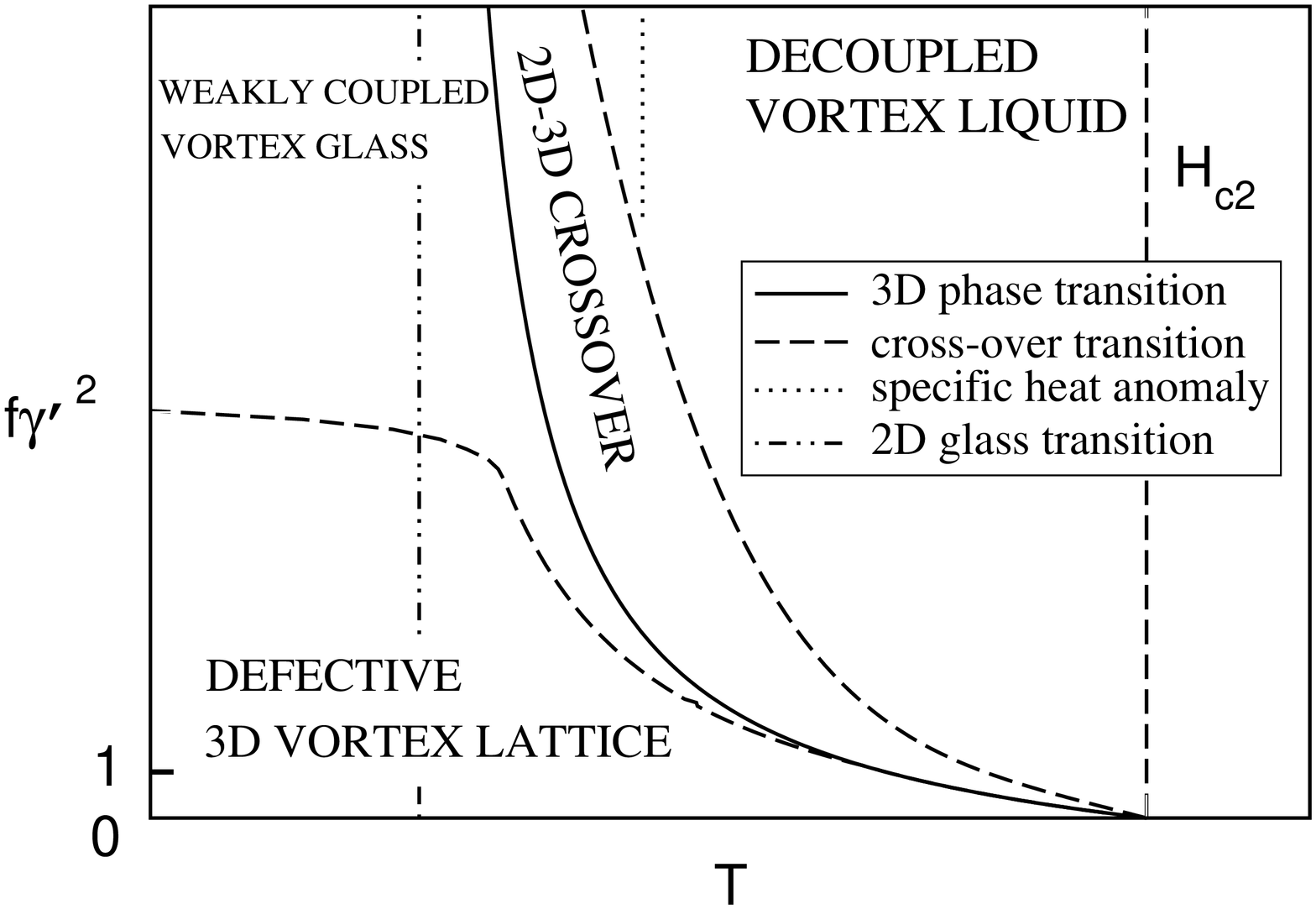}
\caption{Shown is the proposed phase diagram assuming weak point pins
and a continuous vortex glass phase transition for isolated layers.
The concentration of in-plane vortices, $f$, is held fixed,
and a   mean-field temperature dependence,
$J\propto T_{c0} - T$, is assumed.
Monte Carlo simulations of the same $XY$ model studied here
find evidence for a second-order transition  between the
vortex-glass and the vortex-liquid phases
(see ref. \cite{olsson}).
When confronted with the first-order decoupling transition
that is expected to separate the vortex liquid from the 3D vortex lattice
(see refs.  \cite{jpr00} and  \cite{daemen}),
this implies the existence of  a critical endpoint 
consistent with experiments on YBCO (ref. \cite{bouquet}) 
and with other  numerical simulations (ref. \cite{nono1})
of the  present $XY$ model.
A transition to a  3D vortex lattice without defects is reported
in ref. \cite{bg} at $f\gamma^{\prime 2} < 1$.}
\label{phasedia}
\end{figure}

The results of  the above duality method are summerized by
the phase diagram displayed in Fig. \ref{phasedia}.
The present theory notably predicts that
an inverted specific heat anomaly
(jump followed by a tailoff)
occurs at weak coupling
in the vortex liquid
when the 2D correlation length  $\xi_{2D}$ 
matches the 2D disorder scale     $l_{2D}$
if $\xi_{2D}$ diverges faster than $(T - T_g^{(2D)})^{-1}$
[see Eq. (\ref{c_perp})].  
Such a feature has in fact been observed
within the vortex liquid phase of YBCO\cite{bouquet}.
The weight of the latter peak is about
$\Delta e_{\rm exp}\cong 6\, {\rm mJ}/{\rm cm}^3$,
while the peak shown by Eq. (\ref{c_perp}) has a weight
$\Delta e_{J} =  (\Phi_0^2/16\pi^3 \lambda_L^2 \Lambda_0^2) 
 \Delta \overline{\langle{\rm cos}\, \phi_{l, l+1}\rangle}$. 
Equating these and using values of 
$\lambda_L\cong 140\, {\rm nm}$ and $\Lambda_0\cong 7\, {\rm nm}$
for the penetration depths in YBCO\cite{blatter}
yields a $10\,$\% jump in the ``cosine''.




Last, although  recent Monte Carlo simulations
 of the same $XY$ model studied here do indeed
 find evidence for a phase-coherent vortex glass at\cite{nono1}  
$f\gamma^{\prime 2} = 16$ and at\cite{olsson}
$f\gamma^{\prime 2} = 8$,
another one\cite{teitel} using $f\gamma^{\prime 2} = 2$ does not.
We believe that the zero-temperature cross-over
shown in Fig. \ref{phasedia}
 between an entangled vortex glass and a 3D vortex lattice
containing dislocations is
the origin of this discrepancy.

In conclusion, a 
duality analysis of the 
$XY$ model 
finds that random  point pins\cite{larkin}\cite{bulaevskii}
 drive a cross-over
transition in the zero-temperature limit
between defective vortex lattices that show 
strong versus
weak superconductivity across layers
as a function of the Josephson coupling\cite{jpr00}.
We further propose that the inverted specific heat anomaly observed recently
inside of the vortex liquid phase of YBCO 
does not signal a phase transition\cite{bouquet},
but  rather is due to the thermodynamic resonance found here,
Eq. (\ref{c_perp}).

\acknowledgments

The author thanks 
O.  Bernal and  Y. Nonomura
 for discussions.

{\it Note added:}  
Recent Monte Carlo simulations
of the same $XY$ model studied here
also find a non-critical specific-heat anomaly
in the vortex-liquid phase (see ref. \cite{nono2}).

\begin{thebibliography}{}
\bibitem{blatter}
{G. Blatter, M.V. Feigel'man, V.B. Geshkenbein, A.I. Larkin, and V.M. Vinokur,
Rev. Mod. Phys. {\bf 66}, 1125 (1994).}  

\bibitem{friedel}
{J. Friedel, J. Phys. (Paris) {\bf 49}, 1561 (1988).}

\bibitem{korshunov}
{S.E. Korshunov, Europhys. Lett. {\bf 11}, 757 (1990).}

\bibitem{supersolid}
{L.I. Glazman and A.E. Koshelev, Phys. Rev. B {\bf 43}, 2835 (1991);
 M. Feigel'man, V.B. Geshkenbein, and A.I. Larkin,
Physica C {\bf 167}, 177 (1990);
E. Frey, D.R. Nelson, and D.S. Fisher,
Phys. Rev. B {\bf 49}, 9723 (1994).}

\bibitem{ffh}
{D.S. Fisher, M.P.A. Fisher and D.A. Huse, Phys. Rev. B {\bf 43},
130 (1991).}

\bibitem{koshelev97}
{A.E. Koshelev, Phys. Rev. B {\bf 56}, 11201 (1997).}

\bibitem{jpr02}
{J.P. Rodriguez, Phys. Rev. B {\bf 66}, 214506 (2002);
{\bf 69}, 069901(E) (2004).}

\bibitem{cec03}
{C.E. Creffield and J.P. Rodriguez,
Phys. Rev. B {\bf 67}, 144510 (2003).}

\bibitem{larkin} {A.E. Koshelev, L.I. Glazman and A.I. Larkin,
Phys. Rev. B {\bf 53}, 2786 (1996).}

\bibitem{bulaevskii} {L.N. Bulaevskii, M.B. Gaifullin, Y. Matsuda and 
M.P. Maley, Phys. Rev. B {\bf 63}, 140503 (2001).}

\bibitem{jpr00}
{J.P. Rodriguez, Phys. Rev. B {\bf 62}, 9117 (2000).}

\bibitem{n-s}
{C. Zeng, P.L. Leath, and D.S. Fisher,
Phys. Rev. Lett. {\bf 82}, 1935 (1999);
T. Nattermann and S. Scheidl, Adv. Phys. {\bf 49}, 607 (2000).}

\bibitem{bouquet}
{F. Bouquet, C. Marcenat, E. Steep, R. Calemczuk, W.K. Kwok,
U. Welp, G.W. Crabtree, R.A. Fisher, N.E. Phillips and A. Schilling,
Nature {\bf 411}, 448 (2001).}

\bibitem{villain}
{J.V. Jos\' e, L.P. Kadanoff, S. Kirkpatrick and
D.R. Nelson, Phys. Rev. B {\bf 16}, 1217 (1977).}




\bibitem{eps}
{P. Minnhagen and G.G. Warren, Phys. Rev. B {\bf 24}, 2526 (1981).}

\bibitem{jpr01}
{J.P. Rodriguez, Phys. Rev. Lett. {\bf 87}, 207001 (2001).}


\bibitem{mc}
{J.P. Rodriguez and C.E. Creffield, unpublished.}



\bibitem{koshelev96} {A.E. Koshelev, Phys. Rev. Lett. {\bf 77}, 3901 (1996).}


\bibitem{NH}
{D.R. Nelson and B.I. Halperin, Phys. Rev. B {\bf 19}, 2457 (1979).}
     

\bibitem{jpr04}
{J.P. Rodriguez, Physica C {\bf 404}, 311 (2004).}

\bibitem{jpr97}
{J.P. Rodriguez, J. Phys. Cond. Matter {\bf 9}, 5117 (1997).}


\bibitem{olsson} P. Olsson, Phys. Rev. Lett. {\bf 91}, 077002 (2003).


\bibitem{daemen} L.L. Daemen, L.N. Bulaevskii, M.P. Maley and
J.Y. Coulter,  Phys. Rev. B {\bf 47}, 11291 (1993).

\bibitem{nono1}
{Y. Nonomura and X. Hu, Phys. Rev. Lett. {\bf 86}, 5140 (2001).}



\bibitem{bg} {A.E. Koshelev and V.M. Vinokur, 
Phys. Rev. B {\bf 57}, 8026 (1998).}

\bibitem{teitel}
{P. Olsson and S. Teitel, Phys. Rev. Lett. {\bf 87}, 137001 (2001).}

\bibitem{nono2} Y. Nonomura and X. Hu, cond-mat/0302597.







\end{thebibliography}
\end{document}